# On Static and Dynamic Properties of Solitons in Low-Dimensional Systems

Authors: Irina Bariakhtar[1], Victor Bariakhtar[2], Alex Nazarenko[3]
[1]Boston College, USA, [2]Institute of Magnetism, Kyiv, Ukraine, [3]Harvard University, USA

The cross section for scattering of x-rays by solitons is calculated. The authors consider solitons corresponding to the formation of a kink in a system of adatoms on the surface of a substrate, or of a crowdion in a chain of atoms in a crystal that are described by the sine-Gordon equation. It is shown that investigation of the x-ray scattering makes it possible to obtain information about the static and dynamic properties of the solitons.

It is well known that the study of x-ray scattering gives information about the arrangement of atoms in a solid, about the crystal lattice of the solid and about the defects of the crystal lattice [1]. Great attention has been paid recently to special defects in low dimensional systems such as one- and two-dimensional crystals — namely, solitons of various types [2-7]. Solitons, as a rule, are defects which size $x_0$ is considerably greater than the interatomic spacings $a$ and their dimensionality $d_s$ is smaller by one than the dimensionality $d$ of the space of the crystal. The condition $x_0 \gg a$ makes it possible to describe such defects (solitons) macroscopically, while the condition $d_s = d - 1$ permits us to assert that solitons will lead to broadening of x-ray spots.

The aim of this paper is to draw the attention of researchers to the fact the study of the scattering of x-rays (or, incidentally, the scattering of light or electrons), together with an investigation of the neutron scattering, can give important experimental information about the properties of solitons in solids [8].

In this paper we calculate the structure factors of solitons. We consider examples of solitons corresponding to the formation of a kink (fold) in a system of adatoms on the surface of a substrate, or of a crowdion in a chain of atoms in a crystal that are described by the sine-Gordon equation; we also discuss the change of the x-ray line shape in momentum space. It is shown that the temporal structure factor in the expression for the cross section for x-ray scattering coincides with this structure factor in the cross section for neutron scattering.

First, we shall discuss the one-dimensional solitons that arise in two-dimensional (quasi-two-dimensional) crystals. As examples of such crystals we may consider adatoms adsorbed on the surface of a metal or graphite, Si thin films that are used in optoelectronics or PV elements, liquid-crystal films, or lattices of magnetic bubbles in thin magnetic films. Finally, there is a large class of real crystals in which the interaction of atoms between certain crystallographic planes is considerably smaller than the interaction of atoms within one such crystallographic plane; these are quasi-two-dimensional crystals. We recall that mica, high-temperature superconductors, etc., belong to this group.

For definiteness, we shall consider first of all the scattering of x-rays by a soliton corresponding to the sine-Gordon equation. Such solitons arise, e.g., in a system of adatoms interacting weakly with a substrate.

We shall assume that the lattice of adatoms is rhombic and corresponds to a "flat" substrate lattice. We choose the coordinate axes along the symmetry axes of the crystal, and denote the lattice constant along the x axis by a. Let $u(x, y)$ be the displacement of the atoms; then the energy of the two-dimensional crystal (system of adatoms) can be represented in the form

$$\mathcal{H} = \frac{1}{2}\int [\rho_s (\frac{\partial u}{\partial t})^2 + \lambda_{11}u_{xx}^2 + \lambda_{22}u_{yy}^2 + 2\lambda_{33}u_{xy}^2 + \lambda_{12}u_{xx}u_{yy} + 2V_0(1 - \cos(2\pi\tau u))]dxdy \quad (1)$$

where $\rho_s$ is the matter density of the two-dimensional crystal, the $\lambda$ are elastic constants, and $V_0$ is the potential-energy constant. In this formula, the first term is the kinetic energy, the next group of terms is the elastic energy, and the last term describes the interaction of the adatoms with the substrate. The interaction energy is chosen so that it vanishes for the undeformed lattice ($u_x = na$, $u_y = mb$; $n$ and $m$ are integers), i.e., it is reckoned from the energy of adsorption of the atoms; $\tau$ is a reciprocal-lattice vector, and $u_{ik}$ are components of the strain tensor: $2u_{ik} = \frac{\partial u_i}{\partial x_k} + \frac{\partial u_k}{\partial x_i}$.

The Hamiltonian (1) for a displacement $u$ along the $x$ axis corresponds to an equation of motion that is called the sine-Gordon equation:

$$\frac{\partial^2 u}{\partial t^2} - c^2\frac{\partial^2 u}{\partial x^2} + \frac{2\pi V_0}{a\rho_s}\sin\left(\frac{2\pi u}{a}\right) = 0$$

(2)

where $c^2 = \lambda_{11}/\rho_s$.

By now, this equation has been well studied. We give its solution, corresponding to a stationary soliton of the kink type

$$tg\left(\frac{\pi u}{2a}\right) = \exp\left[\mp\frac{x - x_s}{x_0}\right]$$

(3)

where $x_s$ is the coordinate of the center of the soliton, $x_0 = [(a/2\pi)^2 \left(\frac{\lambda_{11}}{V_0}\right)]^{\frac{1}{2}}$ is the width of the soliton, and the $\mp$ signs correspond to a kink and an antikink, respectively. Henceforth, for definiteness, we shall consider a kink. For adatoms interacting weakly with the substrate we have $V_0 \ll \lambda$ and $x_0 \gg a$, i.e., the above macroscopic treatment is justified.
Using Eq. (3), it is not difficult to find the strain in the crystal:

$$u_{xx} = \frac{\partial u}{\partial x} = -\frac{a}{\pi x_0}ch^{-1}(\frac{x-x_s}{x_0}) \ . \quad (4)$$

From this we see that the strain decreases rapidly (exponentially) with distance from the position $x_s$ of the kink. Associated with these strains is a change of the density $n$ of the two-dimensional crystal:

$$n(x) = n_0(x)[1 - u_{xx}(x)] = n_0\left[1 + \frac{a}{\pi x_0}ch^{-1}\left(\frac{x-x_s}{x_0}\right)\right] \qquad (5)$$

where $n_0$ is the equilibrium density of adatoms, or the density of atoms on the surface of the substrate.

We note that both the Hamiltonian (1) and the solution (3) describe a crowdion in a chain of atoms, and Eq.(5) describes the density distribution in the crowdion. We recall that $n_0(x)$ is a periodic function of $x$, with the period $a$ of the lattice. Since $u_{xx} < 0$ ($u_{xx} > 0$), we may take as the reason for the formation of a soliton (3) of the kink (antikink) type the need to place adatoms (vacancies) that are "surplus" in relation to the substrate. To ensure electrical neutrality, the change of density of the ions induces a change of density of the electrons, so that

$$n_e(x) = n_{0e}(x)\left[1 + \frac{a}{\pi x_0}ch^{-1}\left(\frac{x-x_s}{x_0}\right)\right] \qquad (5')$$

Equation (3) describes a stationary soliton. If it moves as a whole with a velocity $v$, the distribution for the displacement vector will be described as before by Eq. (3), in which $x$ must be replaced by $x-vt$ and $x_0$ by $x_0[1 - v/c^2]^{1/2}$.

In this case ($v \neq 0$), from the condition of electrical neutrality and adiabaticity ($\omega_s \ll \Delta\varepsilon_e$, where $\omega_s$ is the characteristic soliton frequency and $\Delta\varepsilon_e$ is the change of the electron energy), knowledge of the ion density makes it possible to determine the change of the electron density. For the characteristic soliton frequencies $\omega_s$ we must take $\omega_s = v/x_s \approx (\partial u/\partial t)/u$. We note that the adiabaticity condition is fulfilled everywhere with the possible exception of a small region of velocities close to the sound velocity $c$.

From the known electron density $n_s(x)$, we can determine the differential cross section for elastic scattering of x-rays [8,9]:

$$d\sigma = 1/2 r_0^2(1 + cos^2\theta)|n_e(\mathbf{q})|^2\delta(\omega - \omega')d\omega'dO',$$

$$n_e(\mathbf{q}, t) = e^{i\mathcal{H}t}n_e(\mathbf{q})t^{-i\mathcal{H}t}, \qquad (6)$$

where $r_0^2 = (e^2/mc^2)$ is the classical electron radius, $\theta$ is the angle between the wave vectors $\mathbf{k}$ and $\mathbf{k}'$ of the incident quantum and the scattered quantum, $\omega$ and $\omega'$ are the frequencies of the incident and scattered quanta, $dO'$ is an element of solid angle in the direction of $\mathbf{k}'$, and $\mathbf{q} = \mathbf{k} - \mathbf{k}'$.

Equation (6) for the scattering of $\gamma$ quanta can be represented in the form

$$d\sigma = \frac{1}{2}r_0^2(1+\cos^2\theta)\int_{-\infty}^{\infty} e^{i\Omega t}dt\,\{Sp\rho n_e(\mathbf{q})n_e(\mathbf{q},t)\}d\omega'dO', \tag{7}$$

where $\rho$ is the density matrix of the crystal, $\Omega = \omega' - \omega$.

$\mathcal{H}$ is the Hamiltonian of the system, and $n_e$ is the electron-density operator. In this form, Eq. (7) describes not only elastic, but also inelastic scattering of x-rays.

Noting that
$$n_e(x) = n_{0e}(x)[1 - u_{xx}(x)],$$

and having in mind the case of scattering with small momentum transfers ($a\mathbf{q} \ll 1$), we rewrite Eq. (7) in the form

$$d\sigma(\Omega, q) = 1/2 r_0^2(1+\cos^2\theta)(n_0 V)^2\{\delta(\omega - \omega')\Delta(k - k') + <u_{xx}(x,0)u_{zz}(x',t)>_{\Omega,q}\}d\omega'dO',$$

$$<u_{xx}(x,0)u_{xx}(x',t)>_{\Omega q} = \frac{1}{V}\int_{-\infty}^{\infty} e^{i\Omega t}\,dt\int d^3x'\exp(-i\mathbf{q}(\mathbf{x}-\mathbf{x}'))<u_{xx}(x,0)u_{xx}(x',t)>_{\Omega,q} \tag{8}$$

where $n_0$ is the uniform electron density and the angular brackets denote both averaging with the density matrix $\rho$ and averaging over the random parameters associated with the solitons; $V$ is the volume.

The first term in Eq.(7) describes ordinary forward scattering by the crystal, and the second term describes scattering by solitons. We shall consider the second term:

$$d\sigma_s(\Omega, q) = 1/2 r_0^2(1+\cos^2\theta)(n_0 V)^2 <u_{xx}(x,0)u_{xx}(x',t)>_{\Omega,q} d\omega'dO' \tag{9}$$

From this we see that the line shape for small-angle scattering (the central peak) of x-rays by a soliton is determined by the correlation function of the tensor of the deformations induced by the soliton. As usual, the running coordinate of the $i$-th soliton will be reckoned from its center $x_{si}$. Then,

$$u_{xx}(x,t) = \sum_{i=1}^{n_s} u_{xx}^{(i)}[x - x_{si}(t)] \tag{10}$$

The summation over $i$ denotes a summation over the solitons, the number of which is $n_s$.

The Fourier component $u_{xx}(\mathbf{q}, t)$ then takes the form:

$$u_{xx}(\mathbf{q}, t) = \frac{1}{V}\int e^{-i\mathbf{q}x}u_{xx}(x,t)d^3x = \sum_{i=1}^{n_s} u_{xx}(\mathbf{q})\exp[i\mathbf{q}\mathbf{x_{si}}(t)]. \tag{11}$$

There are two random parameters characterizing each soliton $i$. One of them is the random initial position $x_{si}(0)$ of the soliton center in the crystal, and the second is related to the random forces (impacts) acting on the soliton as it moves.

The averaging over $x_{si}(0)$ is performed in the standard way:

$$< exp\{iq[x_{si}(0) - x_{si'}(0)]\} >_{x_s} = \Delta(i - i') \qquad (12)$$

Therefore, after the averaging over $x_{si}$, we have for the correlation function

$$< u_{xx}(x, 0)u_{xx}(x', t) >_{\Omega q} = \sum_{i=1}^{n_s} |u_{xx}^{(i)}(q)|^2 \ < \exp[i\mathbf{q}\Delta\mathbf{x_{si}}(t)] >_\Omega . \qquad (13)$$

The angular brackets in the right-hand side of this formula denote averaging over the random motion of the soliton, and $\Delta x_{si}(t) = x_{si}(t) - x_{si}(0)$.

If all the solitons are the same, we have

$$< u_{xx}(x, 0)u_{zz}(x', t) >_{\Omega q} = n_s(T)|u_{xx}(q)|^2 < \exp[i\mathbf{q}\Delta\mathbf{x_s}(t)] >_\Omega, \qquad (14)$$

where $n_s$ is the number of solitons in the body at temperature $T$. From this, it can be seen that the shape of the central peak of the x-ray scattering is determined by the same factor

$$F_t(\Omega, q) = \langle \exp[i\mathbf{q}\Delta\mathbf{x_s}(t)\rangle_\Omega \qquad (15)$$

as for neutron scattering [10].

The intensity of the central x-ray scattering peak, according to (14) and (9), is proportional to the square of the Fourier component of the tensor of the static deformations that are caused by the soliton.

If the motion of the soliton is a random process and is such that all averages of odd powers of $\Delta x_s(t)$ are equal to zero (a Gaussian process), then

$$F_t(\Omega, q) = 2\int_0^\infty exp\left\{-\frac{1}{2}q^2\langle\Delta x_s^2(t)\rangle\right\} cos(\Omega t) dt. \qquad (16)$$

Using Eqs.(14) and (16), we represent the cross section for the small-angle scattering of x-rays in the form:

$$d\sigma_s(\Omega, q) = \frac{1}{2}r_0^2(1 + cos^2\theta)(n_0V)^2 \ n_s(T)u_{xx}^2(q)F_t(\Omega, q). \qquad (17)$$

In this article, we shall consider mainly elastic x-ray scattering, i.e., Eq. (6) will be used. The question of the central peak, i.e., the "smearing out" of $\delta(\omega - \omega')$ into a smoother function as a result of the thermal motion of the solitons, is investigated here by the method proposed in Ref.[10] in a calculation of the magnetic cross section for scattering of neutrons by solitons in a ferromagnet with magnetic anisotropy of the "easy plane" type.

Using (5') for $n_e(x)$, it is easy to find the Fourier component for the electron-density change due to the solitons:

$$n_e(q) = \sum_\tau \exp[-ix_s(q - \tau)] u_{xx}(q - \tau)]n_{0e}(\tau)\Delta(q_y - \tau_y) \qquad (18)$$

In this formula,

$$u_{xx}(q) = \frac{2a}{\pi x_0 ch\left(\frac{\pi x_0 q_x}{2}\right)}. \qquad (19)$$

Using (18) and (19) for $n_e(q)$, it is not difficult to obtain an explicit expression for the cross section for elastic scattering of x-rays by solitons. First, however, we shall discuss (18). From this formula, it can be seen that a soliton with its symmetry axis along the $y$ axis does not give rise to smearing out of the x-ray spot along the $y$ axis, because of the presence of the dependence $\Delta(q_y - \tau_y)$. The situation with the shape of the spot along the $x$ axis is different. Here, we have a superposition of a "point" spot resulting from $n_{0e}(x)$ and a spot "smeared out" to a width $|\tau_x - q_x| \approx 2/\pi x_0$ from the contribution of $u_{xx}(\tau_x - q_x)$. As a result, the x-ray spot is smeared out along the $x$ axis. The magnitude of this smearing is $|\Delta q| \approx 1/x_0 \ll 1/a$, by virtue of the condition $x_0 \gg a$. Therefore, the overlap of individual peaks may be disregarded and the cross-section $d\sigma_s$ for scattering by solitons can be represented in the form of a sum of the partial terms:

$$d\sigma_s = \sum_\tau d\sigma_{s\tau} . \tag{20}$$

Before writing out the expressions for $d\sigma_{s\tau}$ in the explicit form, we note that it is necessary to average the correlator of the densities over the random positions of the coordinate $x_s(0)$ of the soliton center. After the averaging for the part of this correlator due to the solitons, we have

$$< n_e^+(q)\, n_e(q,t) >_\tau = n_s(T)|u_{xx}(\tau_x - q_x)|^2 |n_{0e}(\tau)V|^2 \Delta(\tau_y - q_y) . \tag{21}$$

For $d\sigma_{s\tau}$, in the approximation of stationary solitons, we obtain

$$d\sigma_{s\tau} = \frac{1}{2}r_0^2(1 + \cos^2\theta)n_s(T)|n_{0e}(\tau)V|^2|u_{xx}(\tau_x - q_x)|^2 \Delta(\tau_y - q_y)\delta(\omega - \omega')d\omega' dO' . \tag{22}$$

This formula describes the contribution of the solitons to the elastic scattering of x-rays by solitons distributed randomly along the $x$ axis; the quantity $u_{xx}(\tau_x - q_x)$ is given by the expression (19). For the case when the symmetry axis of the soliton is parallel to the $x$ axis, Eq. (22) can be trivially re-written: $\Delta(\tau_y - q_y)$ is replaced by $\Delta(\tau_x - q_x)$ and $u_{xx}(\tau_x - q_x)$ in the second term is replaced by $u_{yy}(\tau_y - q_y)$.

The thermal motion is taken into account in different ways for a one-dimensional soliton in a system of adatoms and for a crowdion. This is due to the fact that thermal motion for a soliton in a system of adatoms consists in flexural oscillations of the soliton (we recall that the mass of this soliton is proportional to $L_y$, i.e., tends to infinity as $L_y \to \infty$ ($L_y$ is the length of the crystal along the $y$ axis). As regards to the crowdion, its effective mass $m_*$, as is well known, is finite, and under the action of temperature the crowdion executes random motion. In order to obtain the formulas that describe the "smoothing" of the $\delta$-function over the frequencies on account of the thermal motion, we shall consider a crowdion moving with a constant velocity $v$. As already noted, the distribution of the displacement vector for such a soliton will be described by Eq. (3) with the replacements $x \to x - vt$ and $x_0 \to x_0[1 - (v/c)^2]^{1/2}$. Equation (18) for the Fourier component of the change of electron density then takes the form

$$n_e(q,t) = \sum_\tau \exp\left[ix_s(\tau_x - q_x) + i(q_x - \tau_x)vt\right] u_{xx}(\tau_x - q_x)n_{0e}(\tau)\Delta(\tau_y - q_y) . \tag{23}$$

This change in the formula for the Fourier component of the electron density leads to the replacement of $\delta(\omega - \omega')$ in Eq. (22) by $\delta[\omega - \omega' - R(\tau_x - q_x) - v(\tau_x - q_x)]$, where $R(q) = q^2/2m_*$ is the energy of recoil of the crowdion. We shall average this $\delta$-function over the thermal motion of the crowdion:

$$\langle\delta(\Omega - qv)\rangle = \left(\frac{m_*}{2\pi T}\right)^{1/2} \int_0^\infty \delta(\Omega - qv)\exp\left(-\frac{m_*v^2}{2T}\right)dv = \left(\frac{m_*}{2\pi q^2 T}\right)^{1/2} \exp\left(-\frac{m_*\Omega^2}{2q^2 T}\right). \quad (24)$$

Using this formula for the contribution of the solitons to the x-ray scattering, we finally have

$$d\sigma_{s\tau} = \frac{1}{2}r_0^2(1 + \cos^2\theta)n_s(T)|n_{0e}(\tau)V|^2 F_R(\tau_x - q_x)F_t(\omega - \omega', \tau_x - q_x)\Delta(\tau_y - q_y)d\omega' dO', \quad (25)$$

where $F_R$ and $F_t$ are the spatial and temporal structure factors, which are given by the formulae

$$F_R(q) = |u_{xx}(q)|^2 = 2a/\pi x_0 ch\left(\frac{\pi x_0 q_x}{2}\right)$$

$$F_t(\Omega, q) = \left(\frac{m_*}{2\pi q^2 T}\right)^{\frac{1}{2}} \exp\left(-\frac{m_*(\Omega - R)^2}{2q^2 T}\right) \quad . \quad (26)$$

We note that the spatial structure factor $F_R$ is determined by the actual form of the distribution of the displacements of the atoms in the soliton. The temporal structure factor $F_t$, on the other hand, arises as a result of averaging of the $\delta$-function describing the law of energy conservation in the x-ray scattering process, and has a general Gaussian form in the approximation of an ideal gas of solitons.

We shall discuss in more detail the line shape for small-angle x-ray scattering — the line shape of the central peak (CP) of the x-ray scattering. For this we return to Eq. (17), but we can also start from Eq. (25), in which we set $\tau = 0$ and replace $n_{0e}(0)$ by $n_0$. We recall that allowance for the interaction of solitons with each other and with phonons, lattice defects, etc., changes the formula for the neutron-scattering CP from a Gaussian form to a smoother form [11]. As already noted, the CP for x-ray scattering has the same nature as the CP for neutron scattering. The indicated scattering processes change the form of the temporal structure factor in Eq. (17) from a Gaussian [see (26)] to a smoother, Lorentzian form. To calculate this change, as can be seen from Eq.(16), it is necessary to return to the analysis of the character of the soliton motion. As is well known [11, 12], under the action of random collisions, a soliton executes Brownian motion, which is described by two diffusion coefficients $D$ and $D_*$. The kinetic equation for the soliton distribution function $f$ has the form:

$$\frac{\partial f}{\partial t} + v\frac{\partial f}{\partial x} = \frac{T}{D}\frac{\partial}{\partial v}\left[fv + \frac{T}{m_*}\frac{\partial f}{\partial v}\right] + D_*\frac{\partial^2 f}{\partial x^2} \quad . \quad (27)$$

The right-hand side of this equation is the collision integral in the Fokker-Planck approximation and the approximation of small gradients of the distribution function.

Here, $D$ and $D_*$ are the normal and anomalous soliton-diffusion coefficients, $m_*$ is the effective mass of the soliton, and $T$ is the temperature of the thermostat.

Using Eq. (27), we can calculate the average $\langle \Delta x_s^2(t) \rangle$, and with it the structure factor $F_t(\Omega, q)$. We shall give the results for $F_t(\Omega, q)$, following Ref. [13].

If the momentum transfer and frequency transfer are large ($\Omega\tau \gg 1, ql \gg 1$ where $\tau$ and $l$ are the soliton mean free time and mean free path), $F_t(\Omega, q))$ has the form (26).

If $\Omega\tau \ll 1,$ and $ql \ll 1,$ the line shape is replaced by a Lorentzian one:
$$F_t(\Omega, q) = Dq^2/[\Omega^2 + (Dq^2)^2]. \qquad (28)$$
In the region of intermediate frequencies, the anomalous diffusion of solitons ($D_*$) can occur.

Investigation of the scattering of x-rays by a soliton makes it possible to determine the dependence of the soliton density on the temperature from the factor $n_s(T)$, the shape of the particle-density distribution in the soliton from a study of the shape of the spot in momentum space, i.e., from the shape of $F_R(q)$, and the soliton mass from a study of $F_t(\Omega, q)$.

In summary, the authors developed a new approach to describing the electronic density of the low-dimensional systems with solitons based on the sine-Gordon equation of motion, and constructed Hamiltonian for such systems. The equation for the x-ray scattering by solitons with the additional term to the original Bragg scattering due to the solitons was obtained theoretically. Also, the form-factors that are related to the solitons' movement and interaction are constructed. It was suggested that by comparing the experimental results to the presented theory, one can determine the existence of the solitons in a system under consideration.